\documentclass[
10pt,
aps,
twocolumn,
groupedaddress,
superscriptaddress,
amsmath,
amssymb,
pra,
nofootinbib,
longbibliography]{revtex4-2}
\usepackage{mathtools,amsmath,amsxtra}
\usepackage{amssymb}
\usepackage{bbm}
\usepackage{physics}
\usepackage[english]{babel}
\usepackage{dsfont}
\usepackage{graphicx}
\usepackage{dcolumn}
\usepackage{bm}
\usepackage{bbold}
\usepackage{bigints}
\usepackage[usenames,dvipsnames]{xcolor}
\usepackage{cancel}
\usepackage{tikz}
\usetikzlibrary{decorations.pathmorphing}
\usetikzlibrary{arrows.meta}
\usepackage{bm}
\usepackage{braket}
\usepackage[normalem]{ulem}
\usepackage{lipsum}
\usepackage{subcaption}
\usepackage{mathrsfs}
\usepackage[colorlinks,citecolor=Blue,linkcolor=Red, urlcolor=Blue]{hyperref}
\definecolor{REVISIONCOLOR}{RGB}{120,45,145}

\begin{document}

\title{Optimal probes for estimating the dephasing of a quantum spin system}

\author{Leonardo Ratini}
\email{leonardo.ratini@unipr.it}
\affiliation{Universit\`a di Parma, Dipartimento di Scienze Matematiche, Fisiche e Informatiche, I-43124 Parma, Italy} 
\affiliation{INFN–Sezione di Milano-Bicocca, gruppo collegato di Parma, 43124 Parma, Italy}
\author{Stefano Carretta}
\email{stefano.carretta@unipr.it}
\affiliation{Universit\`a di Parma, Dipartimento di Scienze Matematiche, Fisiche e Informatiche, I-43124 Parma, Italy}
\affiliation{INFN–Sezione di Milano-Bicocca, gruppo collegato di Parma, 43124 Parma, Italy}
\affiliation{Consorzio Interuniversitario Nazionale per la Scienza e Tecnologia dei Materiali (INSTM), I-50121 Firenze, Italy}
\author{Francesco Albarelli}
\email{francesco.albarelli@unipr.it}
\affiliation{Universit\`a di Parma, Dipartimento di Scienze Matematiche, Fisiche e Informatiche, I-43124 Parma, Italy} 
\affiliation{INFN–Sezione di Milano-Bicocca, gruppo collegato di Parma, 43124 Parma, Italy}

\begin{abstract}
We investigate the optimal estimation of dephasing in finite-dimensional quantum spin systems.
We first discuss some properties of the optimal probe states and present an efficient numerical algorithm to compute the optimal precision in dephasing estimation and identify the optimal probe states.
For the dephasing channel arising from Gaussian phase diffusion, a spin cat state is known to be optimal for small dephasing.
We analytically determine the range of dephasing strengths over which the cat state remains globally optimal for spins $J=1$ and $J=3/2$, and derive upper bounds on its optimality region for larger spins.
We then compare numerically several probe states proposed in the literature, showing in particular that the simple class of probes obtained from uniform superpositions of all levels performs near-optimally for stronger dephasing.
Finally, we apply our framework to a central-spin model coupled to a nuclear-spin bath, whose dynamics we simulate using the cluster-correlation expansion.
Our results show that, while the cat state is optimal for weak Gaussian dephasing, pulse-controlled non-Markovian dynamics can favor states with amplitudes distributed over multiple spin levels, and in some cases the uniform state outperforms the other fixed probes considered.
This shows how the choice of probe state depends both on the dephasing strength and on the dynamics.
\end{abstract}

\maketitle

\twocolumngrid 

\section{Introduction}

Coherence is one of the defining features of quantum systems, formally represented by off-diagonal elements of the density matrix in a reference basis~\cite{Streltsov2017a}.
Usually it is a fragile property and inevitable interaction with an environment, either quantum or classical, suppresses quantum coherence.
A standard case is when only the off-diagonal elements are affected by the noise, while the diagonal ones, i.e. the populations, are preserved; this is what we generally call \emph{pure dephasing}.
In spin systems, coherence decay is commonly characterised by the transverse relaxation time $T_2$, pure dephasing is the regime in which population relaxation can be neglected \cite{Slichter1990,Degen2017}.
Understanding this process is important for operating quantum devices, to characterise the noise and to develop strategies for mitigating its effects~\cite{Szankowski2017a,Petiziol2021,Ratini2025}.

While quantum metrology has traditionally focused on estimating parameters related to the coherent part of the evolution, such as frequency, external fields or phase, the precise estimation of noise parameters also has important applications~\cite{Demkowicz-Dobrzanski2015a,Pirandola2017a}.
In both settings, optimising the initial probe state and the measurement is essential to determine the optimal precision achievable with the available resources.
Noise affects the choice of the optimal probe state, and for coherent parameters it strongly limits the advantage given by quantum resources such as entanglement or squeezing~\cite{Huelga1997,Demkowicz-Dobrzanski2017}.
However, when the noise itself depends on the parameter of interest, the loss of coherence can be used to extract information about the environment.
More specifically, dephasing can be exploited to characterise noise spectra~\cite{Alvarez2011,Benedetti2014a,Benedetti2018,Razavian2019a,Norris2016,Szankowski2017a,Wang2021h,Sung2019}, estimate environmental coupling strengths and correlation times~\cite{Benedetti2014d,Zwick2016a}, and perform quantum thermometry~\cite{Razavian2019,Gebbia2020,Mitchison2020,Wang2021h,Brattegard2024,Yu2024a,Brattegard2026}.

In this context, much work has been devoted to single-qubit probes~\cite{Degen2017,Benedetti2014a,Zwick2016a}, but the advantages of multiqubit probes have also been investigated~\cite{Rossi2015,Gebbia2020,Wang2026o,Brady2026a,Brady2026,Prabhu2026,Paz-Silva2017,Wang2026c}.
Higher-spin systems, i.e. qudits, have similarly been considered for environmental characterisation through dephasing~\cite{Knysh2013,Candeloro2021,Albarelli2023}.
Pulse-controlled qudit spectroscopy has also been developed to reconstruct spectra of stationary Gaussian dephasing noise~\cite{Javaherian2025a}.
For a single qubit, an equal superposition of the two states of the reference basis is optimal, while
for a larger spin and for multiple qubits, the initial state can instead distribute its populations over several levels, and a question arises on how to choose the optimal superposition of levels.
In a prototypical model, sometimes simply known as phase diffusion~\cite{Knysh2013}, the dephasing is modeled as a random phase encoded by a spin operator and the phase has Gaussian statistics.
In this case, coherences between widely separated levels are more sensitive to weak dephasing, but they are also suppressed more rapidly as the dephasing strength increases.
This makes choosing the optimal probe state a nontrivial problem.

For dephasing channels that damp all coherences by the same factor, uniform superpositions of the reference basis states have been proven optimal for estimating the dephasing strength~\cite{Yadin2019}.
For phase diffusion, Knysh and Durkin~\cite{Knysh2013} have shown that in the large-spin limit the amplitudes of an asymptotically optimal probe follow a cosine profile.
In the bosonic setting, the optimal estimation precision in the absence of energy constraints has been derived and can be approached by uniform superpositions of an increasing number of Fock states~\cite{Huang2024a}.
More recently, frequency estimation of stochastic AC fields was connected to time-dependent collective dephasing of multiple qubits~\cite{Dey2025a}.
Inspired by the earlier bosonic results, this study investigated uniform superpositions of Dicke states as multiqubit probes.

These results motivate a more detailed characterisation of optimal probes for dephasing estimation at finite spin and dephasing strength.
In this paper we study this question by first deriving structural properties of optimal probes for general dephasing channels and formulating a convex optimisation that yields both an optimal probe and its achievable precision, quantified by the quantum Fisher information (QFI).

We then apply this framework to Gaussian phase diffusion generated by $J_z$.
In this model, a natural probe for weak dephasing is an equal superposition of the extremal spin states, known as a spin cat state and equivalent to a multiqubit Greenberger--Horne--Zeilinger (GHZ) state
\footnote{The correspondence holds once the fully symmetric subspace of $N$ qubits is identified with a spin-$J=N/2$ system.
For collective dephasing, restricting the probe to this subspace does not reduce the precision; see Appendix~\ref{app:spin_mapping}.}.
Such probes are particularly attractive because both their preparation and an optimal measurement can be chosen independently of the unknown dephasing strength.
Previous investigations have explored the conditions under which these probes are optimal~\cite{Rossi2015,Albarelli2023}, but their range of global optimality at finite spin has not been fully characterised.
We analytically determine the exact range of dephasing strengths over which the spin cat state is globally optimal for $J=1$ and $J=3/2$, and derive upper bounds on this range for larger spins.
To quantify the precision that can be gained by populating additional levels, we complement these analytical results with numerical comparisons of representative classes of states, including uniform superpositions.
These comparisons show that a simple one-parameter family interpolating between cat and uniform probes closely approaches the optimum over a wide range of spins and dephasing strengths.

We also consider Gaussian dephasing generated by $J_z^2$, for which the conventional spin cat is insensitive to the noise and useful probes have a different population structure.
This example illustrates how the spectrum of the noise operator, together with the dephasing strength, determines the optimal levels to populate.

Finally, we study a more realistic setting, modeling a typical solid-state spin coupled to a nuclear-spin bath, simulating the dynamics with the cluster correlation expansion (CCE) technique \cite{Yang2008,Yang2009a}, and comparing probe states under free and pulse-controlled evolution.
Overall, our results provide analytical criteria and numerical tools for selecting probes that extract information from dephasing in multilevel quantum sensors.

The methods developed here are relevant to multilevel quantum platforms subject to dephasing, including neutral-atom qudits and solid-state impurity spins, such as antimony atoms in silicon.

\section{Preliminaries}

\subsection{Quantum estimation theory}
\label{subsec:estimation_preliminaries}

We consider a single real parameter $\theta$ encoded in the output $\rho_\theta=\Lambda_\theta(\rho)$ of a quantum channel $\Lambda_\theta$, formally this is a parametric family of completely-positive trace-preserving (CPTP) maps. 
The input state $\rho$ does not depend on the parameter $\theta$ and has the role of a probe to be optimized to obtain a good precision in estimating the value of $\theta$.
A measurement is described as a positive operator-valued measure (POVM) $\{M_x\}$, with $\sum_x M_x=  \mathbbm{1}$; the outcome probabilities are given by the Born rule $p_{\theta}(x)=\operatorname{Tr}(\rho_\theta M_x)$.
The classical Fisher information
\begin{equation}
    F_C[p_\theta]=\sum_{x\, ; \, p_{\theta}(x)>0}
    \frac{[\partial_\theta p_{\theta}(x)]^2}{p_{\theta}(x)},
\end{equation}
quantifies the asymptotically achievable precision.
More formally, for $\nu$ independent repetitions, the variance of a locally unbiased estimator $\tilde\theta$ obeys the classical and quantum Cram\'er--Rao bounds~\cite{Demkowicz-Dobrzanski2015a},
\begin{equation}
    \operatorname{Var}(\tilde\theta)\geq\frac{1}{\nu F_C[p_\theta]}
    \geq\frac{1}{\nu F_Q[\rho_\theta]}.
    \label{eq:prelim_qcrb}
\end{equation}
Independent repetitions can correspond to measurements performed sequentially on identical preparations of the system, but also to simultaneous measurements on an ensemble of independently prepared probes undergoing independent noise realizations. 
Indeed, the QFI $F_Q$ is the maximum of $F_C$ over measurements and can be expressed as
\begin{equation}
    F_Q[\rho_\theta]=\operatorname{Tr}(\rho_\theta L_\theta^2),
    \qquad 2\partial_\theta\rho_\theta=L_\theta\rho_\theta+\rho_\theta L_\theta,
    \label{eq:prelim_sld}
\end{equation}
where $L_\theta$ is a hermitian operator known as symmetric logarithmic derivative (SLD); a projective measurement on its eigenvectors attains $F_C = F_Q$.
Under standard regularity conditions, both bounds in Eq.~\eqref{eq:prelim_qcrb} are asymptotically attainable as $\nu$ grows.
However, since the optimal probe and measurement may depend on the operating value of $\theta$, this optimization of estimation is local.

An important property of the QFI is its monotonicity under parameter-independent CPTP maps $\mathcal E$, i.e.   
\begin{equation}
    F_Q[\mathcal E(\rho_\theta)]\leq F_Q[\rho_\theta].
    \label{eq:qfi_monotonicity}
\end{equation}
This property is a form of data processing inequality and means that processing the state before measuring, it cannot provide more information than performing the optimal measurement directly on the original state.
Equality holds for unitary transformations and isometries, which preserve all the information encoded in the state.

The largest output QFI attainable by optimising over input states, including those entangled with an auxiliary noiseless reference system, is known as \emph{channel QFI} and denoted as $F_Q[\Lambda_\theta]$.
We will see that for dephasing channels this optimum can always be attained by the probe system alone, without the need for a reference system.
Optimising over probe states is generally difficult and closed-form expressions of the channel QFI are known only in special cases.
However, allowing the probe state to be entangled with an auxiliary noiseless system leads to a semidefinite program (SDP) to compute the channel QFI~\cite{Fujiwara2008,Demkowicz-Dobrzanski2015a,Zhou2021a}.
For a differentiable Kraus representation of the channel $\Lambda_\theta(\rho)=\sum_k K_k\rho K_k^\dagger$, we introduce an operator $\alpha$
\begin{equation}
\begin{aligned}
    \dot K_k(h)&=\partial_\theta K_k-i\sum_l h_{kl}K_l,\qquad h=h^\dagger,\\
    \alpha_\theta(h)&=\sum_k\dot K_k(h)^\dagger\dot K_k(h) \;, 
\end{aligned}
    \label{eq:prelim_alpha}
\end{equation}
where the Hermitian matrix $h$ describes parameter-dependent changes of Kraus representation, which leaves the channel unchanged but can affect the parametric derivatives of the Kraus operators.
The channel QFI  can then be expressed as
\begin{equation}
\begin{aligned}
    F_{Q}[\Lambda_\theta]
    &:=\max_{\rho_{SA}}F_Q[(\Lambda_\theta\otimes\mathcal I_A)(\rho_{SA})]\\
    &=4\min_{h=h^\dagger}\|\alpha_\theta(h)\|,
\end{aligned}
    \label{eq:prelim_channel_qfi}
\end{equation}
where $\|\cdot\|$ is the operator norm, i.e. the maximum singular value of the matrix.
This expression can explicitly be recast as an SDP by standard manipulations~\cite{Demkowicz-Dobrzanski2015a}.

\subsection{Dephasing channels}
\label{subsec:dephasing_preliminaries}

We consider an abstract finite-dimensional quantum system of dimension $d$, i.e. a qudit.
Once an orthonormal basis $\{\ket m\}_{m=1}^d$ is fixed, a choice usually dictated by the physics of the problem, a dephasing channel\footnote{These maps are also known as \emph{generalized} dephasing channels, a subclass of so-called Hadamard channels~\cite{Bradler2010}.} is a map that preserves populations and multiplies each coherence by a factor:
\begin{equation}
    \Lambda_\theta(\rho)=E_\theta\circ\rho,
    \qquad E_\theta\succeq0,\qquad (E_\theta)_{mm}=1.
    \label{eq:general_schur_family}
\end{equation}
Here $\circ$ denotes the so-called Hadamard or entrywise product.
Positive semidefiniteness of the matrix $E_\theta$, denoted by the symbol $\succeq$, and the fact that its diagonal elements are all equal to unity ensure complete positivity and trace preservation of the channel, respectively.
We call the $E_\theta$ the dephasing matrix, and generally it may be complex and need not possess any further symmetry, beyond hermitianity and positive semidefiniteness.

We stress that a qudit can always be understood as a spin-$J$ quantum system, and we will mostly use this terminology in this paper.
For a spin $J$ with $d=2J+1$, we consider the spin operator $J_z$ along the $z$ direction; its eigenstates $J_z\ket m=m\ket m$ ($m=-J,\ldots,J$ and $\hbar=1$) are chosen as the reference basis with respect to which coherence is defined.

A family of dephasing channels is obtained as the average over random phase rotations distributed with a probability density $w_\theta(\phi)$:
\begin{equation}
    \Lambda_\theta(\rho)=\int_{\mathbb R}d\phi\,
    w_\theta(\phi)e^{-i\phi J_z}\rho e^{i\phi J_z},
\end{equation}
and the corresponding dephasing matrix reads
\begin{equation}
    (E_\theta)_{mn} = \int_{\mathbb R}d\phi\,w_\theta(\phi)e^{-i (m-n) \phi}.
\end{equation}
Such dephasing channels, obtained by spin rotation, give rise to a matrix $E_\theta$ that is Toeplitz, i.e. its entries depend only on the difference of the level indices.
If one further assumes an even phase distribution $w_{\theta}(\phi)=w_{\theta}(-\phi)$, the dephasing matrix becomes also real and symmetric, invariant under reflection $m\mapsto-m$.

Our main example of dephasing channel is Gaussian phase diffusion~\cite{Knysh2013}, obtained from a Gaussian phase distribution with variance $2\Delta$ and zero-mean:
\begin{equation}
    w_\Delta(\phi)=\frac{e^{-\phi^2/(4\Delta)}}{\sqrt{4\pi\Delta}},
    \qquad (E_\Delta)_{mn}=e^{-\Delta(m-n)^2}.
    \label{eq:prelim_gaussian_diffusion}
\end{equation}
In the following, the parameter to be estimated will be the dephasing strength $\Delta>0$, and we will assume that the mean phase is known or irrelevant.
This channel also includes Markovian pure-dephasing dynamics
\begin{equation}
    \frac{d\rho}{dt}=-\gamma[J_z,[J_z,\rho]],
    \qquad\Delta=\gamma t,
    \label{eq:prelim_lindblad}
\end{equation}
where the Lindblad jump operator is $\sqrt{2\gamma}\,J_z$ for $\gamma\geq0$.

Here $t$ is the so-called interrogation time, which in many cases can be chosen in the experiment.
If the parameter of interest is the dephasing rate $\gamma$, its QFI is obtained by the expression for the QFI under reparametrisations~\cite{Paris2009} $F_\gamma(t)=t^2F_\Delta|_{\Delta=\gamma t}$, where $F_\Delta$ denotes the QFI for estimating $\Delta$ with the same probe.
The Gaussian-channel results in Sec.~\ref{sec:Gaussian-phase-diffusion} are presented in terms of the accumulated dephasing $\Delta$, instead of the dephasing rate and interrogation time.

We stress that, when comparing different interrogation times, it is customary to also account for the number of repetitions.
For a fixed total interrogation time $T$, we can perform $\nu\simeq T/t$ independent runs, ideally neglecting preparation and readout times.
\footnote{
Neglecting these additional times is an idealization.
If each run requires a fixed additional time $t_{\mathrm d}$ for preparation, readout and, when needed, reinitialization of the environment, the relevant quantity at fixed total experimental time is instead $F_\theta(t)/(t+t_{\mathrm d})$.
This can change the optimal interrogation time and the relative advantage of different probes~\cite{Dooley2016}.}
Writing $F_\theta(t)$ for the output QFI of a single interrogation, the total accumulated QFI is
\begin{equation}
    F_\theta^{\mathrm{tot}}=\nu F_\theta(t)
    \simeq T\frac{F_\theta(t)}{t}.
    \label{eq:qfi_time_budget}
\end{equation}
Thus, by Eq.~\eqref{eq:prelim_qcrb}, the relevant quantity to maximize at fixed $T$ is the QFI rate $F_\theta(t)/t$.

For estimating a Markovian dephasing rate, the optimal QFI rate is approached as $t\to0$, by rapidly repeating probe preparation, a short evolution and measurement~\cite{Gardner2025b,Wang2026c}.
For the dynamics in Eq.~\eqref{eq:prelim_lindblad}, a spin cat state maximizes the variance of $J_z$ and achieves this short-time limit.
In Sec.~\ref{subsec:cat_optimality} we will discuss the region of optimality of spin cat probes, which also identifies the finite interrogation times for which it remains the best probe.

The matrix $E_\Delta$ in Eq.~\eqref{eq:prelim_gaussian_diffusion} is Toeplitz, due to the equally spaced spectrum of the generator $J_z$ that imprints the phase.
If we replace $J_z$ by a generator $G=\sum_m g_m\ket m\bra m$ diagonal in the same basis, we obtain the dephasing matrix $(E_\theta)_{mn}=\int_{\mathbb R}d\phi\,w_\theta(\phi)e^{-i (g_m-g_n) \phi}$.
For unequally spaced $g_m$ this matrix is not Toeplitz.
Focusing again on Gaussian fluctuations, we obtain $(E_\Delta)_{mn}=e^{-\Delta(g_m-g_n)^2}$.
More concretely, in Sec.~\ref{subsec:quadratic_diffusion} we will consider a quadratic generator $G=J_z^2$ as a simple example that features relevant qualitative differences from the case of a linear generator.

For a random-phase channel with a fixed generator $G$, a simple upper bound on the channel QFI is given by the classical Fisher information of the distribution~\cite{Dey2025a,Huang2024a}:
\begin{equation}
    F_Q[\Lambda_\theta]\leq F_C[w_\theta]
    =\int_{\mathbb R}d\phi\,
    \frac{[\partial_\theta w_\theta(\phi)]^2}{w_\theta(\phi)}.
    \label{eq:classical_phase_bound}
\end{equation}
Since the phase distribution contains all the parameter dependence, measuring the probe cannot give more information than an ideal observation of the random phase itself, this technique is also known as a classical simulation of the channel~\cite{Demkowicz-Dobrzanski2015a,Pirandola2017a}.
For the Gaussian distribution in Eq.~\eqref{eq:prelim_gaussian_diffusion}, we obtain the bound
\begin{equation}
    F_Q[\Lambda_\Delta]\leq\frac{1}{2\Delta^2}.
    \label{eq:gaussian_phase_bound}
\end{equation}
This bound is independent of the spin and holds for both the linear and quadratic generators considered below, valid whenever the dephasing matrix comes from a random phase.

\section{Optimal probes for dephasing estimation}
\subsection{Populations of the probe states and concavity}
\label{subsec:probe_populations}
We now start exploring the properties of optimal probe states for the general case of a generic differentiable parametric family of dephasing channels of the form in Eq.~\eqref{eq:general_schur_family}, for which the matrix $E_\theta$ may be complex and need not be Toeplitz.

By convexity of the QFI in the input state~\cite{Demkowicz-Dobrzanski2015a}, we can restrict to pure probe states.
Their phases are irrelevant, since a diagonal and parameter-independent unitary commutes with the channel and leaves the output QFI unchanged.
Thus, we parametrise probe states in terms of the probability vector $p$, i.e. the populations
\begin{equation}
\begin{aligned}
    \ket{\psi_p}&=\sum_{m=1}^{d}\sqrt{p_m}\ket{m},
    &p_m&\geq0,\quad\sum_m p_m=1,\\
    \rho_\theta(p)&=\Lambda_\theta(\ket{\psi_p}\bra{\psi_p}),
    &F_\theta(p)&=F_Q[\rho_\theta(p)].
\end{aligned}
    \label{eq:population_probe}
\end{equation}

It is easy to show that entanglement with an auxiliary system $A$ does not improve the QFI.
Indeed, any pure input state can be expressed as
\begin{equation}
    \ket{\Psi}=\sum_m\sqrt{p_m}\ket{m}\ket{a_m},
\end{equation}
where the conditional states of the auxiliary system are normalized, but not necessarily orthogonal.
The parameter-independent map $V\ket{m}=\ket{m}\ket{a_m}$ is an isometry, and we have
\begin{equation}
    (\Lambda_\theta\otimes\mathcal I_A)(\ket{\Psi}\bra{\Psi})
    =V\rho_\theta(p)V^\dagger.
    \label{eq:schur_no_ancilla}
\end{equation}
Thus, the two output families have the same QFI, and again convexity also excludes any potential improvement from mixed joint input states.
Hence $F_{Q}[\Lambda_{\theta}]$ is attained without the need for an auxiliary noiseless system.

A key property for this specific problem is that $F_\theta(p)$ is \emph{concave in the probe populations}\footnote{For clarity, we point out that averaging the populations and preparing the corresponding pure superposition is different from mixing the two probe states; convexity of the QFI in the input state concerns the second operation.}
\begin{equation}
    F_\theta\bigl(\lambda p+(1-\lambda)q\bigr)
    \geq\lambda F_\theta(p)+(1-\lambda)F_\theta(q), 
    \label{eq:population_concavity}
\end{equation}
with $0<\lambda<1$.
This can be shown by considering a map with the following diagonal Kraus operators
\begin{equation}
\begin{aligned}
    D_1&=\sum_{m}\sqrt{\frac{\lambda p_m}{\lambda p_m +(1-\lambda)q_m }}\ket{m}\bra{m} ,\\
    D_2&=\sum_{m}\sqrt{\frac{(1-\lambda)q_m}{\lambda p_m +(1-\lambda)q_m }}\ket{m}\bra{m}.
\end{aligned}
\end{equation}
the sum is restricted to $\lambda p_m +(1-\lambda)q_m >0$ and they satisfy $D_1^\dagger D_1+D_2^\dagger D_2= \mathbbm{1}$ on this support.
Interpreting these operators as a measurement with outcomes $1$ and $2$, we can consider the case in which the outcome is saved in an orthogonal classical register.
Formally this means considering the rectangular Kraus operators $|1\rangle_R\otimes D_1$ and $|2\rangle_R\otimes D_2$, where we have introduced orthonormal register states $\langle i|j\rangle_R=\delta_{ij}$.
This defines a parameter-independent channel with a classical outcome register that acts as
\begin{align}
    \rho_\theta\bigl(\lambda p+(1-\lambda)q\bigr)
    &\longmapsto \lambda |1\rangle\langle1|_R\otimes\rho_\theta(p) \nonumber\\
    &\quad +(1-\lambda)|2\rangle\langle2|_R\otimes\rho_\theta(q) \nonumber\\
    &=\lambda\rho_\theta(p)\oplus(1-\lambda)\rho_\theta(q).
\end{align}
Thanks to the monotonicity of the QFI, we obtain Eq.~\eqref{eq:population_concavity}, since the QFI of a classical-quantum state with parameter-independent weights, in this case $\lambda$ and $1-\lambda$, is just the weighted sum of the QFI of each block.

\subsection{Reflection symmetry and optimal probes}

Now suppose that the channel family is invariant under reversal of the levels $\ket{m} \mapsto \ket{d+1-m}$.
Formally this means that the channel commutes with the action of the unitary $U={\sum_{m=1}^d}\ket{d+1-m}\bra{m}$:
\begin{equation}
    \Lambda_\theta(U\rho U^\dagger)
    =U\Lambda_\theta(\rho)U^\dagger,
    \label{channel_sym}
\end{equation}
which in turn means that the dephasing matrix is invariant as 
\begin{equation}
    U E_\theta U^\dagger=E_\theta \; .
    \label{comm_EU}
\end{equation}
We assume that this holds for all the relevant values of the parameter $\theta$.
While a real symmetric Toeplitz dephasing matrix satisfies this condition, reflection invariance is a more general property.

By considering the reversed populations $(p^{\mathrm{rev}})_m=p_{d+1-m}$, we thus immediately obtain $F_\theta(p^{\mathrm{rev}})=F_\theta(p)$.
However, thanks to the concavity property we have just shown, we have
\begin{equation}
    F_\theta\!\left(\frac{p+p^{\mathrm{rev}}}{2}\right)
    \geq F_\theta(p).
    \label{eq:symmetric_probe_populations}
\end{equation}
This means that the optimization can be restricted to pure probes with real, non-negative, reflection-symmetric amplitudes.
We stress that it is the \emph{population vector} that is symmetrized, not the density matrix of the input state.

We now consider a family of populations $p(\epsilon) = p(0) + \epsilon p_1$, i.e. an affine function of $\epsilon$.
Since we have seen that the function $F_\theta(p(\epsilon))$ is concave, its right derivative at $\epsilon=0$ satisfies
\begin{equation}
    F_\theta(p(\epsilon)) \leq F_\theta(p(0)) + \epsilon \frac{d }{d \epsilon } F_\theta(p(\epsilon)) \big\vert_{\epsilon = 0}\qquad 0\leq\epsilon\leq1. 
    \label{eq:concave_endpoint_test}
\end{equation}
Consequently, $\frac{d }{d \epsilon } F_\theta(p(\epsilon)) \big\vert_{\epsilon = 0} \leq 0$
is necessary and sufficient for $p(0)$ to be a global maximizer of the QFI within this family of populations.
Because of concavity, a nonpositive initial slope cannot increase later, whereas if slope is positive one immediately finds a probe state with a higher QFI.

For $J=1$ and $J=3/2$, the symmetric populations have only one free parameter, so testing the slope at $\epsilon=0$, i.e. a balanced superposition of $\ket{\pm J}$, establishes its global optimality over all probes.
At larger spins, testing a single direction in the space of populations can be used to rule out the optimality of the spin cat state.

\subsection{Numerical algorithm for channel QFI and optimal probes}
\label{subsec:numerical_optimization}
The general structure of a dephasing channel also simplifies the computation of optimal probes beyond analytically treatable cases.
Every Kraus operator can be chosen to be diagonal, $K_k=\sum_m c_{km}\ket m\bra m$.
Introducing the vectors $c_m=(c_{1m},\ldots,c_{rm})^{\mathsf T}$, where $r$ is the number of Kraus operators and $\dot c_m=\partial_\theta c_m$, the operator $\alpha_\theta(h)$ in Eq.~\eqref{eq:prelim_alpha} is also diagonal with entries
\begin{equation}
    [\alpha_\theta(h)]_{mm}=\|\dot c_m-ihc_m\|_2^2 \, ,
    \label{eq:dephasing_alpha_entries}
\end{equation}
where $\|v\|_2^2=\sum_j|v_j|^2$ is the squared Euclidean vector norm.
Therefore the matrix constraint of the general SDP reduces to one bound on a vector norm for each level. Equivalently, the channel QFI is obtained as $F_{Q}[\Lambda_{\theta}]=4{t^\star}^2$ where $t^\star$ solves the optimisation
\begin{equation}
\begin{aligned}
    &\underset{t\in\mathbb R,\ h=h^\dagger}{\operatorname{minimise}}\quad t,\\
    &\text{subject to}\quad \|\dot c_m-ihc_m\|_2\leq t,
    \quad m=1,\ldots,d,
\end{aligned}
    \label{eq:schur_socp}
\end{equation}
Each vector is affine in the optimisation variables, so this is a \emph{second-order cone program} (SOCP)~\cite{Lobo1998,Boyd2004}.
This formulation replaces a general matrix positivity constraint by separate norm constraints; the computational gain depends on the dimension and on the specific solver.
In Appendix~\ref{app:socp} we present the details of this derivation.

Interestingly, solving this optimisation also yields an optimal probe that attains the channel QFI. 
Indeed, once the optimal variables $t^\star$ and $h^\star$ have been obtained, an optimal probe population vector can be recovered from
\begin{equation}
\begin{aligned} 
& \sum_{m:p_m>0}p_m \nabla_h\sqrt{[\alpha_\theta(h)]_{mm}}\vert_{h = h^\star}=0
 \\ & \text{subject to}\quad p_m= 0 \quad  \text{if} \quad [\alpha_\theta(h^\star)]_{mm}< {t^\star}^2.
\end{aligned}
\end{equation}
In practice, one does not need to solve this problem directly.
Conic solvers return a dual variable $\mu_m^\star\geq0$ for each norm constraint in Eq.~\eqref{eq:schur_socp}, 
together with $t^\star$ and $h^\star$.
These dual variables satisfy $\sum_m\mu_m^\star=1$, and $p_m=\mu_m^\star$ are the populations of an optimal probe.
A single numerical solve therefore gives both the channel QFI and an optimal probe state.
In Appendix~\ref{app:socp}, we derive this result and further explain reductions for dephasing channels with a real, reflection-invariant dephasing matrix.
It provides an efficient numerical procedure for the models considered below.

\section{Gaussian phase diffusion}
\label{sec:Gaussian-phase-diffusion}

In this section we present analytical and numerical results concerning the model of Gaussian phase diffusion, defined in Eq.~\eqref{eq:prelim_gaussian_diffusion}.

\subsection{Optimality of the cat state}
\label{subsec:cat_optimality}
The dephasing matrix for Gaussian phase diffusion has elements $(E_\Delta)_{mn}=e^{-\Delta(m-n)^2}$ and thus has a reflection symmetry. 
The argument outlined in Sec.~\ref{subsec:probe_populations} based on the concavity of the QFI in the probe populations provides an immediate test of cat state optimality.
For $J=1$ and $J=3/2$, every symmetric probe state belongs to a one-parameter family whose extremal point is the spin cat state $(\ket{J}+\ket{-J})/\sqrt{2}$.
The global optimality of the cat state is therefore determined by the sign of the right derivative of the QFI at this point.
More detailed calculations are reported in Appendix~\ref{app:gaussian_derivations}.

For $J=1$, the symmetric probe states are
\begin{equation}
    \ket{\psi_\epsilon}
    =\sqrt{\frac{1-\epsilon}{2}}(\ket{-1}+\ket{1})
    +\sqrt{\epsilon}\ket{0},\qquad 0\leq\epsilon\leq1.
    \label{eq:spin_one_probe}
\end{equation}
The populations are affine in $\epsilon$, and for $\epsilon=0$ we obtain the cat state.

For compactness, we introduce the parameter $u=e^{-2\Delta}\in(0,1)$ and denote the QFI of each family by $f_J(\epsilon)=F_Q[\Lambda_\Delta(\ket{\psi_\epsilon}\bra{\psi_\epsilon})]$.
The derivative at $\epsilon = 0$ is
\begin{equation}
    f'_{1}(0)=-\frac{8uC(u)}{(1-u^2)(1+u^2)^2},
    \label{eq:spin_one_endpoint}
\end{equation}
where
\begin{equation}
    C(u)=2u^5+u^4+4u^3-2u-1.
    \label{eq:spin_one_threshold_polynomial}
\end{equation}
By Eq.~\eqref{eq:concave_endpoint_test}, the cat is globally optimal if and only if $C(u)\geq0$. Descartes' rule of signs gives exactly one positive root; since $C(0)=-1$ and $C(1)=4$, it lies in $(0,1)$. Writing this root as $u_1\approx0.7522$, we obtain
\begin{equation}
    0<\Delta\leq\Delta_1,
    \qquad \Delta_1=-\frac12\log u_1\approx0.1423.
    \label{eq:spin_one_threshold}
\end{equation}
Above this threshold the optimal probe has a nonzero central population. Its explicit value and uniqueness are derived in Appendix~\ref{app:spin_one_qfi}.

\begin{figure*}[!ht]
    \centering
    \includegraphics{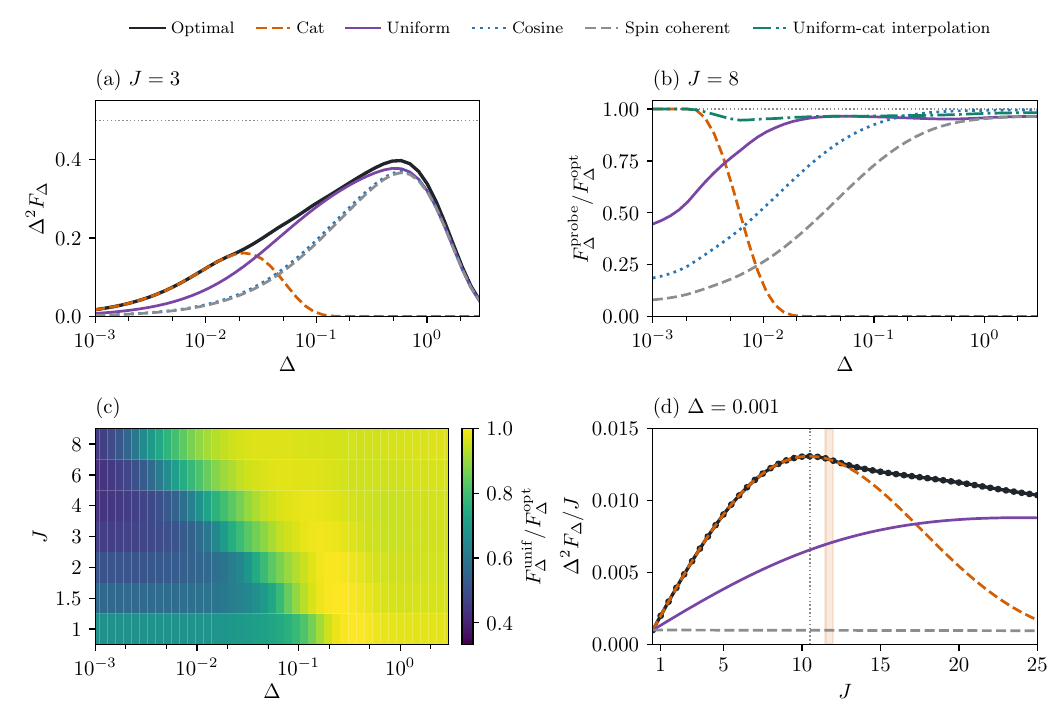}
    \caption{Comparison of the performance for different probe states in estimating the dephasing $\Delta$ for Gaussian phase diffusion generated by $J_z$.
    (a) Normalized QFI for $J=3$; the horizontal dotted line is the classical Gaussian bound $\Delta^2F_\Delta=1/2$ of Eq.~\eqref{eq:gaussian_phase_bound}.
    (b) Efficiency relative to the optimal QFI for $J=8$.
    (c) Uniform-probe efficiency on a grid of $(J,\Delta)$ values.
    (d) Information per $J$ at $\Delta=10^{-3}$ for $J=1/2,1,\ldots,25$.}
    \label{fig:gaussian-dephasing}
\end{figure*}

For $J=3/2$, the corresponding family is
\begin{equation}
\begin{aligned}
    \ket{\psi_\epsilon}
    ={}&\sqrt{\frac{1-\epsilon}{2}}
    (\ket{-3/2}+\ket{3/2})\\
    &+\sqrt{\frac{\epsilon}{2}}
    (\ket{-1/2}+\ket{1/2}),\qquad 0\leq\epsilon\leq1.
\end{aligned}
    \label{eq:spin_three_halves_probe}
\end{equation}
The derivative at the cat endpoint reads
\begin{equation}
    f'_{3/2}(0)=r(u)\frac{t_0(u)}{q_0(u)^2},
    \label{eq:spin_three_halves_endpoint}
\end{equation}
where $r(u)=u/[(u+1)(u^3-1)]<0$, $q_0(u)=\sum_{k=0}^{8}u^k>0$, and the polynomial $t_0$ is given in Appendix~\ref{app:spin_three_halves_qfi}. Concavity makes $t_0(u)\geq0$ necessary and sufficient for global cat optimality. Its coefficients have one sign change, while $t_0(0)=-5$ and $t_0(1)=3888$, so it has exactly one positive root, $u_{3/2}\approx0.878142$.
Therefore, the cat is globally optimal when
\begin{equation}
    0<\Delta\leq\delta_{3/2},
    \qquad \delta_{3/2}=-\frac12\log u_{3/2}\approx0.065.
    \label{eq:spin_three_halves_threshold}
\end{equation}
For both spins, a positive derivative above the threshold proves that populating the inner levels improves the QFI. 

For larger spins, there is more than one independent population parameter after imposing reflection symmetry, and these low-spin results can only give upper bounds on the spin cat optimality region.
We define $\Delta_{\mathrm{cat}}(J)$ as the maximum value of dephasing for which the spin cat is optimal.
For integer $J$, restricting the probe to $\{\ket{-J},\ket{0},\ket{J}\}$ reproduces the spin-one problem at $\Delta_{\mathrm{eff}}=J^2\Delta$.
For $J=3n/2$ with odd positive $n$, the four levels $\{\ket{-3n/2},\ket{-n/2},\ket{n/2},\ket{3n/2}\}$ similarly reproduce the spin-$3/2$ problem at $\Delta_{\mathrm{eff}}=n^2\Delta$.
The associated QFI acquires a common parameter-rescaling factor, which leaves the comparison between probes unchanged.

For the remaining half-integer spins, a bound follows by comparing the spin cat with the superpositions $\ket{\phi_m}=(\ket{-m}+\ket{m})/\sqrt2$ for $0<m<J$, which we call \emph{spin kitten states}.
For each kitten the QFI is obtained easily, since the state remains in a two-dimensional subspace of the whole Hilbert space; we obtain
\begin{equation}
    F_Q[\Lambda_\Delta(\ket{\phi_m}\bra{\phi_m})]
    =\frac{16m^4}{e^{8m^2\Delta}-1}.
    \label{eq:gaussian_cat_qfi}
\end{equation}
The same expression for $m=J$ gives the spin cat QFI.

Choosing $m=J-1$ gives a bound valid for every half-integer $J\geq5/2$, as shown in Appendix~\ref{app:larger_spin_bounds}. Combining these comparisons gives, for $J\geq1$,
\begin{equation}
    \Delta_{\mathrm{cat}}(J)\leq\frac{1}{J^2}
    \begin{cases}
        \Delta_1\approx0.1424, & J\in\mathbb N,\\
        \Delta_{3/2}\approx0.1462, & J=3n/2,\ n\text{ odd},\\
        \Delta_{5/2}\approx0.3143, & \text{other half-integer }J,
    \end{cases}
    \label{eq:cat_region_bounds}
\end{equation}
where $\Delta_{3/2}=9\delta_{3/2}/4$.
These upper bounds $\Delta_{\mathrm{cat}}(J)$ show that retaining the spin cat state as an optimal probe requires dephasing to decrease at least as $J^{-2}$.
This poses stringent constraints on the use of spin cat states as the size of the system increases.

\subsection{Numerical comparison of paradigmatic probe states}

We compare the optimal probes with four representative classes of states: the spin cat, a uniform superposition of all spin levels~\cite{Dey2025a}, a spin-coherent state pointing perpendicular to $z$, and a state with a cosine profile~\cite{Knysh2013}.
All these classes can be characterised in terms of the populations, as shown in Eq.~\eqref{eq:population_probe}.
The spin cat has $p_{\pm J}=1/2$, the uniform superposition $p_m^{\mathrm{unif}}= 1/(2J+1)$, the coherent state $p_m^{\mathrm{coh}}=2^{-2J}\binom{2J}{J+m}$, and the cosine $p_m^{\mathrm{cos}}=\cos^2[\pi m/(2J+1)]/{\mathcal N_J}$ with $\mathcal N_J=\sum_{m=-J}^J\cos^2[\pi m/(2J+1)]$.
Notice that the spin coherent state corresponds to a product state when the spin $J$ is made out of $2J$ qubits, as considered in e.g.~\cite{Rossi2015,Dey2025a}.

We further introduce a family of states that interpolates between cat and uniform probes, by adjusting the total population of the two extremal spin levels while keeping all interior populations equal, formally:
\begin{equation}
\begin{aligned}
    p_{\pm J}(w)&=\frac{1-w}{2},\\
    p_m(w)&=\frac{w}{2J-1},\qquad -J<m<J \; ,
\end{aligned}
\label{eq:one_parameter_probe}
\end{equation}
with $J\geq1$ and $0\leq w\leq1$.
We refer to this family as the uniform-cat interpolation.
The cat state corresponds to $w=0$, while the uniform superposition is recovered for $w=(2J-1)/(2J+1)$.
Since these populations are affine functions of $w$, the concavity established in Sec.~\ref{subsec:probe_populations} reduces the optimization within the family to a one-variable concave maximization.
For $J=1$ and $J=3/2$, the family contains every reflection-symmetric population distribution and therefore attains the full optimum, previously discussed.
In general, the weight $w$ is different for each value of $\Delta$.

In Fig.~\ref{fig:gaussian-dephasing} we compare these probes with the channel QFI obtained using the method introduced in Sec.~\ref{subsec:numerical_optimization}.
The label $F_\Delta$ represents the output QFI of each probe and we denote the channel QFI as $F_\Delta^{\mathrm{opt}}=F_Q[\Lambda_\Delta]$ to explicitly highlight that it corresponds to the optimal probes.
In panels (a) and (d), we also renormalise the QFI with the square of the parameter value, obtaining a dimensionless quantity that measures sensitivity to relative changes of the dephasing strength, also known as quantum signal to noise ratio~\cite{Paris2009}.
The horizontal dotted line in panel (a) is the classical Gaussian bound $\Delta^2F_\Delta\leq1/2$ from Eq.~\eqref{eq:gaussian_phase_bound}.

Panels (b) and (c) quantify the loss of precision due to using a certain probe rather than the optimal one.
We can see that the spin cat is optimal for weak dephasing, but its coherence between the extremal levels is rapidly suppressed as $\Delta$ increases and so its performance quickly degrades. 
Uniform probes contain coherences between levels at all separations. As the dephasing strength increases, their coherences between nearby levels decay more slowly than the extremal coherence of the cat state, and the uniform probes become nearly optimal once the cat state is no longer optimal.
Panel (c) also shows more generally that they are appreciably less efficient in the weak-dephasing region where the spin cat is optimal.
On the other hand, the uniform-cat interpolation family in panel (b) bridges the two regimes by tuning the balance between the extremal and interior populations.
Notice that, although the relative efficiency plotted in panels (b) and (c) is high, this does not correspond to a large absolute QFI, which generally decreases for strong dephasing.

Finally, panel (d) shows the benefit of increasing $J$ for a fixed dephasing strength, $\Delta=10^{-3}$.
Essentially in this plot, the spin number $J$ is considered as a resource, and $F_\Delta/J$ can be interpreted as information per unit resource.
In other words, by maximising this figure of merit, one identifies the optimal spin $J$ to use in each shot. Using a larger $J$ is not beneficial.
This interpretation is especially meaningful when considering the spin $J$ as $2J$ qubits.
The dotted line marks the maximum of the optimized information per spin, for $J=10.5$.
The shaded interval separates $J=11.5$, where the cat agrees with the numerical optimum, from $J=12$, where the discrepancy between the cat and the optimum can be resolved numerically.
This shows that the spin value where the information per spin is maximal and the value where the cat ceases to be optimal are close but distinct.
This implies that increasing the spin can become a less efficient strategy, even while the cat is still numerically optimal.

\begin{figure*}[!ht]
    \centering
    \includegraphics{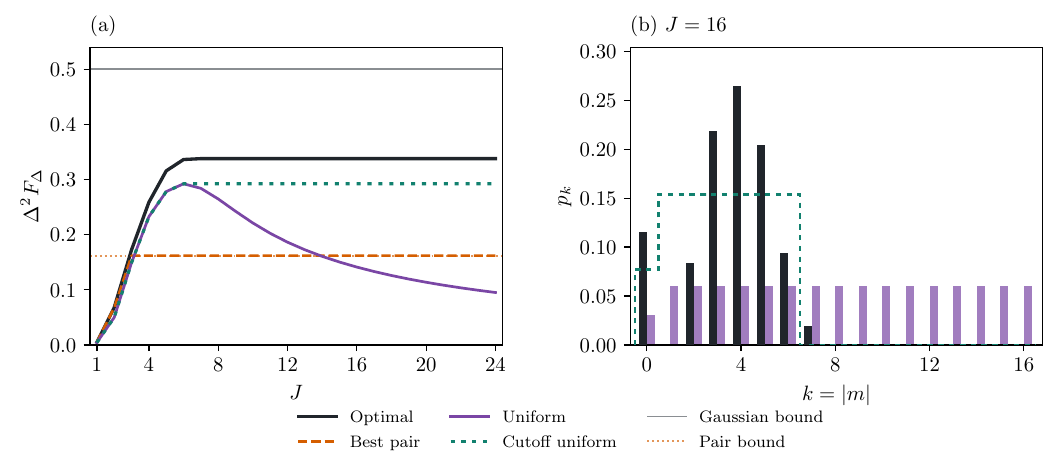}
    \caption{Comparison of probe states for dephasing estimation with Gaussian phase diffusion generated by $J_z^2$, at fixed $\Delta=0.01$.
    (a) Normalized QFI versus integer spin $J$, comparing the optimum, 
    the best two-level probe, the uniform superposition of all levels, and the uniform superposition restricted to $|m|\leq\min(J,6)$.
    (b) Total populations of the eigenspaces $k=|m|$ at $J=16$, with the same colors.}
    \label{fig:gaussian-quadratic-dephasing}
\end{figure*}

\subsection{Phase diffusion with a quadratic generator}

\label{subsec:quadratic_diffusion}

To illustrate how the coupling spectrum affects probe selection, in this section we consider dephasing induced by random fluctuations of a parameter $\phi$ coupled to $J_z^2$, represented by the random unitary $U_\phi=e^{-i\phi J_z^2}$.
Such a coupling can arise from fluctuations of an axial anisotropy parameter.
In atomic spin qudits, a tensor light shift can also produce an energy proportional to the square of the magnetic quantum number, as demonstrated in $^{87}\mathrm{Sr}$~\cite{Ahmed2025a}.
Fluctuations of its coefficient provide another possible realization of this quadratic coupling.
As done previously, we take $\phi$ to be a zero-mean Gaussian random phase with variance $2\Delta$, so that
\begin{equation}
    (E_\Delta)_{mn}=\exp[-\Delta(m^2-n^2)^2].
    \label{eq:quadratic_dephasing}
\end{equation}
The dephasing matrix is no longer Toeplitz but is still reflection symmetric. The parameter to be estimated is again the dephasing strength $\Delta$.

For integer $J$, the distinct eigenvalues of $J_z^2$ are $g_k=k^2$, $k=0,\ldots,J$.
The degeneracy between $\ket{k}$ and $\ket{-k}$ allows the optimization to be performed on just these $J+1$ eigenspaces, and only their total populations matter.
In particular, for this model the spin cat $(\ket{J}+\ket{-J})/\sqrt{2}$ is unaffected by the noise and has zero QFI.
However, other two-level probes can be useful;
within each pair, the QFI is proportional to $p(1-p)$, where $p$ and $1-p$ are the populations of the two eigenspaces, and is therefore maximized by the balanced superposition $p=1/2$.
For a balanced superposition of two distinct eigenspaces with gap $g=|k^2-l^2|$ we obtain
\begin{equation}
    F_\Delta^{(k,l)}=\frac{g^4}{e^{2\Delta g^2}-1}.
    \label{eq:quadratic_pair_qfi}
\end{equation}
We compare the best pair, the numerical optimum and the uniform superposition of all $2J+1$ spin states.

Fig.~\ref{fig:gaussian-quadratic-dephasing}(a) shows that as $J$ increases, it becomes beneficial to use more than two levels; the advantage in using multilevel probes exceeds a factor of two over the best pair of levels.
Higher levels have increasingly large gaps and contribute little information for a fixed dephasing strength; therefore, distributing population over them can reduce performance.
Thus, for the specific value of dephasing in the figure, as the number of spins is increased, the optimised populations are concentrated around $k\leq7$, and adding more level does not improve the performance.
For this reason the uniform superposition has a decreasing performance as $J$ is increased.
Instead, one can consider a cutoff of the uniform probe, also shown in the figure, i.e. a uniform superposition of the spin levels with $|m|\leq\min(J,6)$.
Its support stops growing beyond $J=6$, avoiding the loss of precision caused by spreading population over increasingly separated levels.

Even in this case, the Gaussian bound in Eq.~\eqref{eq:gaussian_phase_bound} applies; however, the maximum observed value $\Delta^2F_\Delta\simeq0.338$ remains below the $1/2$ bound.
Overall, this example shows that multilevel optimization can be useful even when the conventional spin cat is completely insensitive, and that the performance of uniform probes strongly depends on the spectrum of the noise coupling.

\section{Dephasing estimation in a realistic solid-state spin system}

In this section, we focus on spin systems in which the probe consists of one or more spins, while dephasing arises from interactions between the probe and the surrounding spin bath.
Nitrogen vacancy (NV) centers and molecular nanomagnets (MNMs) are representative real-world examples of this class of systems~\cite{Jerger2023,Petiziol2021,Ratini2025}.
The Hamiltonian of the whole system is
\begin{equation}
    H = H_P + H_B + H_{PB}
\end{equation}
Here 
\begin{equation}
H_P=\bm{B}\cdot\bm{\Gamma}\cdot\bm{J} + \bm{J}\cdot\bm{D}\cdot\bm{J},
\label{eq:Hp}
\end{equation}
is the probe Hamiltonian, $\bm{B}$ is the external magnetic field, $\bm{\Gamma}$ is the gyromagnetic ratio tensor, $\bm{D}$ is the self-interaction tensor and $\bm{J}$ is the vector of the spin operators. In the bath Hamiltonian
\begin{equation}
    H_B=\sum_i \bm{B}\cdot\bm{\gamma}^i\cdot\bm{I}_i
    \label{eq:Hb} + \sum_{i,j}\bm{I}_i\cdot \bm{J}^{ij}\cdot\bm{I}_j .
\end{equation}
$\bm{\gamma}^i$ and $\bm{I}_i$ represent the gyromagnetic ratio tensor and the spin operator vector for the $i$th bath spin, respectively. $\bm{J}^{ij}$ is the interaction tensor for bath spins associated with the spin pair $(i,j)$.
Finally, the interaction Hamiltonian
\begin{equation}
    H_{PB}=\sum_{i} \bm{J}\cdot \bm{A}^{i}\cdot \bm{I}_{i}
\end{equation}
contains the interaction tensor $\bm{A}^{i}$ between the probe and the $i$th bath spin. A dipole-dipole interaction is assumed for $\bm{J}^{ij}$ and $\bm{A}^{i}$, then
\begin{equation}
    \bm{J}^{ij}=\frac{\mu_0}{4\pi\abs{\bm{R}_{ij}}^3}\big(\bm{\gamma}^i\cdot \bm{\gamma}^j-3\frac{\bm{\gamma}^i\cdot \bm{R}_{ij}\otimes \bm{\gamma}^j\cdot\bm{R}_{ij}}{\abs{\bm{R}_{ij}}^2}\big), \qquad i\ne j,
\end{equation}
and
\begin{equation}
    \bm{A}^{i}=\frac{\mu_0}{4\pi\abs{\bm{R}_{i}}^3}\big(\bm{\Gamma}\cdot \bm{\gamma}^i-3\frac{\bm{\Gamma}\cdot\bm{R}_{i}\otimes \bm{\gamma}^i\cdot\bm{R}_{i}}{\abs{\bm{R}_{i}}^2}\big),
\end{equation} 
where $\bm{R}_{i}$ ($\bm{R}_{ij}$) is the distance between the $i$th bath spin and the probe ($j$th bath spin). 

Evaluating the relevant figures of merit requires knowledge of the probe density matrix.
However, since the bath typically contains tens or hundreds of spins, direct diagonalization of the full Hamiltonian is computationally infeasible.
To overcome this limitation, we employ the CCE technique.

\subsection{Cluster correlation expansion technique}
\label{sub_sub_sec:CCE}
We start by defining a central-spin system that plays the role of the probe. The central system interacts with the surrounding spin bath and the resulting Hamiltonian is diagonal in the central-spin Hilbert space. If it is not diagonal, a perturbative approach, like the Schrieffer-Wolff transformation~\cite{Slichter1990,Bravyi2011}, can be used to get a diagonal Hamiltonian.
The resulting pure dephasing dynamics can be simulated with the CCE technique that approximates the coherence factor by truncating its expansion up to a certain order. 
Initialize the probe into the superposition
\begin{equation}
    \ket{\psi}_{\alpha, \beta}=C_{\alpha}\ket{\alpha}+C_{\beta}\ket{\beta}
    \label{eq:system_state}
\end{equation}
where $\ket{\alpha}$ and $\ket{\beta}$ are two eigenstates of the central system with coefficients $C_{\alpha}$ and $C_{\beta}$ respectively.
By considering a certain cluster $C$ in the spin bath, the matrix element $(\alpha,\beta)$ of the coherence factor for this cluster will be
\begin{equation}
    L_C^{\alpha \beta}(t)=\operatorname{Tr}_B\big(U_C^{\beta}(t)^{\dagger}U_C^{\alpha}(t) \rho_B \big)
\end{equation}
where $\rho_B$ is the initial density matrix of the bath. 
$U_C^{\alpha}$ ($U_C^{\beta}$) is the time propagator for cluster $C$, conditioned on the central system being in the eigenstate $\ket{\alpha}$ ($\ket{\beta}$). It is generated by the corresponding conditional cluster Hamiltonian, in which the spin operators $\bm{I}_i$ with $i\notin C$ are replaced by their mean-field values.
For the cluster $C$ we define the irreducible coherence factor iteratively as
\begin{equation}
   \widetilde{L}_C^{\alpha \beta}(t)=\frac{L_C^{\alpha\beta}(t)}{\prod_{C^{\prime}\subset C}\widetilde{L}_{C^{\prime}}^{\alpha\beta}(t)} 
   \label{eq:irre_choerence_factor}
\end{equation}
where $C^{\prime}$ runs over the proper subclusters of $C$.
Then, running over each possible cluster $C$ included in the bath, the coherence factor can be expanded as
\begin{equation}
    L^{\alpha \beta}(t)=\prod_C \widetilde{L}_C^{\alpha \beta}(t)
    \label{eq:cce}
\end{equation}
Truncating the expansion at order $N$, we include only the clusters $C$ such that $|C|\le N$. This technique has been used to compute the coherence factor decay in molecular nanomagnets and NV-center systems \cite{Troiani2008,Petiziol2021,Onizhuk2021,Ratini2025}.
The corresponding coherence factor matrix defines the pure-dephasing quantum channel associated with that bath for each time $t$, then $E(t)\equiv L(t)$. The initial density matrix $\rho_P$ of the probe evolves as
\begin{equation}
    \rho_P(t)=E(t)\circ \rho_P
    \label{eq:hadamard_channel}
\end{equation}

To compute $E(t)$ when pulses are performed during the time evolution, we used the generalized CCE (gCCE) developed in~\cite{Yang2020f}.
This technique has been used to investigate MNMs~\cite{Ryan2025} and silicon vacancies~\cite{Fazio2024a} systems when the Hamiltonian is not diagonal in the central system Hilbert space. In our case, we consider the gCCE framework because the pulses applied to the probe affect all density matrix elements simultaneously, rendering the conventional CCE approach physically inconsistent. In gCCE, for cluster $C$ we define
\begin{equation}
    L_C^{\alpha \beta}(t)=\frac{\operatorname{Tr_B}\big(\bra{\alpha}U_C(t)\rho U_C(t)^{\dagger}\ket{\beta} \big)}{\rho_P^{\alpha\beta}}
    \label{eq:coherence_factor_gCCE}
\end{equation}
where $\rho=\rho_P \otimes \rho_B$ is the initial density matrix of the whole system (probe and bath).
Here, $U_C$ is the time evolution operator for cluster $C$, no longer conditioned on the state of the probe as above.
Eqs.~\eqref{eq:irre_choerence_factor} and~\eqref{eq:cce} can be used to find the factor $L^{\alpha \beta}(t)$ that is a function of each component of $\rho_P$.
It follows that Eq.~\eqref{eq:hadamard_channel} is no longer valid because, in general, we do not have a pure-dephasing channel.
For typical spin systems we consider, $L^{\alpha \beta}(t)$ is essentially independent of the initial probe state $\rho_P$ and $L^{\alpha \alpha}(t)\approx 1 \quad \forall t$.
It follows that Eq.~\eqref{eq:hadamard_channel} is still approximately valid with $E(t)\equiv L(t)$.

\subsection{QFI rate for spin-$3/2$ probes}
We choose $J=3/2$ and $\bm{B}=B\hat{\bm z}$, with $B=0.1\,\mathrm{T}$ and $\bm{\Gamma}$ diagonal with $\Gamma_{ii}=2.0\,\mu_B$ for $i=x,y,z$, where $\mu_B$ is Bohr’s magneton. The zero-field splitting tensor is zero, except for the component $D_{zz}=6\,\mathrm{MHz}$. As a prototypical bath, we randomly distributed 50 Hydrogen nuclear spins into a sphere of radius $10\,\text{\AA}$ centered in the spin probe. The minimum distance among each spin pair is $4\,\text{\AA}$. Suppose we can vary each distance $\bm{R}_{i}$ by a factor $\epsilon$, then the $\bm{R}_{i}\rightarrow \epsilon \bm{R}_{i}$. Now, to estimate the value of $\epsilon$ at any time $t$, we compute the QFI for each initial probe state of interest. We choose $\epsilon=1$ as a reference point and compute the map $E_\epsilon(t)$ and its derivative $\left.\partial_\epsilon E_\epsilon(t)\right|_{\epsilon=1}$ by using the CCE technique, as explained in Sec.~\ref{sub_sub_sec:CCE}.

We denote the output state as $\rho_{P,\epsilon}(t)=\Lambda_{\epsilon,t}(\rho_P)=E_\epsilon(t)\circ\rho_P$ and its QFI as $F_\epsilon(t)=F_Q[\rho_{P,\epsilon}(t)]$.
The optimal value is the channel QFI $F_\epsilon^{\mathrm{opt}}(t)=F_Q[\Lambda_{\epsilon,t}]$.
All quantities are evaluated for $\epsilon=1$.

To compare the information obtained with the same total interrogation time, we plot $F_\epsilon(t)/t$, following Eq.~\eqref{eq:qfi_time_budget}.
At fixed $t$, dividing by $t$ does not change which probe is optimal; however, the normalization matters when comparing different durations.
Thus, the peaks of these curves identify the most useful interrogation times for this resource constraint, see the discussion around Eq.~\eqref{eq:qfi_time_budget}.
We assume independent repetitions with the same initial bath state and neglect preparation, readout and bath reinitialization times.

\begin{figure}[t!]
    \includegraphics{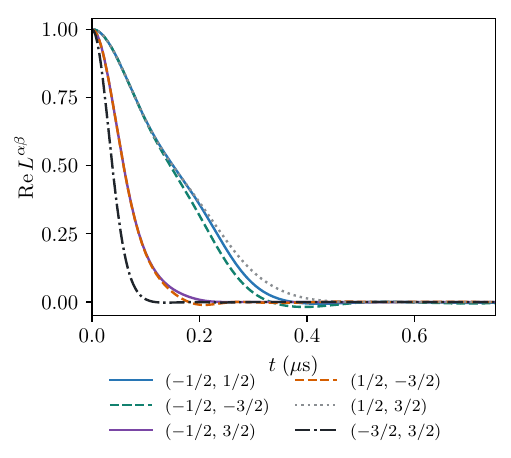}
    \caption{Real part of the coherence factor over time for each pairwise superposition of states $\ket{\alpha}$ and $\ket{\beta}$.}
    \label{fig:FID_coherence}
\end{figure}

In Fig.~\ref{fig:FID_coherence} we report the real part of the coherence factor as a function of time for each pair superposition of states $\ket{\alpha}$ and $\ket{\beta}$. For each time $t$, the values of these curves give the real parts of the off-diagonal elements of the dephasing matrix $E_\epsilon(t)$. As can be deduced from this figure, this map approximates a Toeplitz channel because $(E_\epsilon(t))_{12}\approx (E_\epsilon(t))_{23}\approx (E_\epsilon(t))_{34}$ and $(E_\epsilon(t))_{13}\approx (E_\epsilon(t))_{24}$ for each time $t$.

\begin{figure}[h]
    \includegraphics{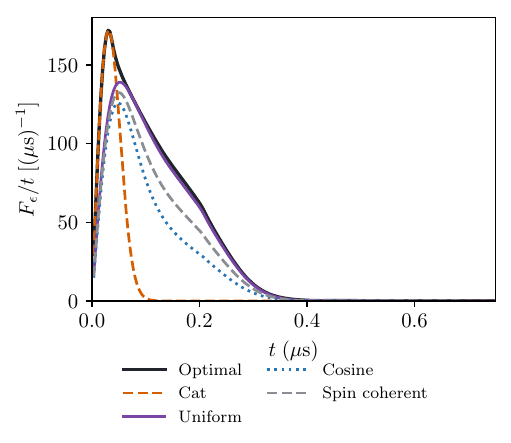}
    \caption{Comparison of the $F_\epsilon(t)/t$ figure of merit for different states in the freely evolving system, with $\epsilon=1.0$.}
    \label{fig:FID_QFIovert}
\end{figure}

Fig.~\ref{fig:FID_QFIovert} reports the figure of merit $F_\epsilon(t)/t$ for different initial states at $\epsilon=1$. Within the parameter range in which the cat state is optimal, it achieves the largest value.

Now we introduce a set of pulses during the free evolution, we want to reproduce a dynamical decoupling scheme that generalizes the echo sequence in such a way that each eigenstate $\ket{\alpha}$ with $\alpha=1,2,3,4$ accumulates the same phase during the evolution. The resulting time evolution operator is
\begin{equation}
    U(t)=e^{-iH\tau}V_1 e^{-iH\tau}V_2 e^{-iH\tau}V_1 e^{-iH\tau}
    \label{eq:ECHO_time_evolution}
\end{equation}
where $H$ is the Hamiltonian of the whole system (probe and bath), $t=4\tau$ and $\tau$ is the free-evolution time between two consecutive pulses.
The denominator in $F_\epsilon(t)/t$ is therefore the duration of the complete sequence.
Here $V_1$ and $V_2$ are the following instantaneous pulses 
\begin{equation}
    V_1= \begin{bmatrix} 
    0 & 0 & 1 & 0 \\ 
    0 & 0 & 0 & 1 \\ 
    1 & 0 & 0 & 0 \\ 
    0 & 1 & 0 & 0 
    \end{bmatrix} \qquad 
    V_2= \begin{bmatrix} 
    0 & 1 & 0 & 0 \\ 
    1 & 0 & 0 & 0 \\ 
    0 & 0 & 0 & 1 \\ 
    0 & 0 & 1 & 0 
    \end{bmatrix}
\end{equation}
The propagator for each cluster $C$ appearing in Eq.~\eqref{eq:coherence_factor_gCCE} can be obtained by field-averaging the spin operators $\bm{I}_i$ in $H$ for $i\notin C$.

\begin{figure}[]
    \includegraphics{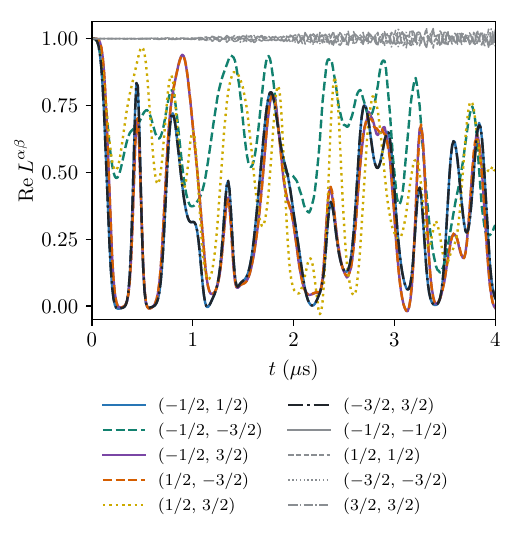}
    \caption{Real parts of the coherence-factor matrix elements over time. Note that $L^{\alpha \alpha}\approx 1$ for $\alpha=1,2,3,4$. Here the total evolution time range is four times the free-evolution time $\tau$, as can be seen in Eq.~\eqref{eq:ECHO_time_evolution}.}
    \label{fig:ECHO_coherence}
\end{figure}

The coherence factor curves for this time evolution are reported in Fig.~\ref{fig:ECHO_coherence}. Note that the gCCE technique has been used here to obtain all the elements of the dephasing matrix $E_\epsilon(t)$. This means that the diagonal elements of $E_\epsilon(t)$ can get values different from 1 but, as can be seen in Fig.~\ref{fig:ECHO_coherence}, $(E_\epsilon(t))_{\alpha\alpha}\approx 1$ for $\alpha=1,2,3,4$. Moreover, the curves do not change if we vary the initial state of the probe, then we deduce that this dynamics approximately reproduces a pure-dephasing channel, as described in Sec.~\ref{sub_sub_sec:CCE}. Compared to Fig.~\ref{fig:FID_coherence}, in Fig.~\ref{fig:ECHO_coherence} the dynamics is dominated by coherence revivals. 

\begin{figure}[h]
    \includegraphics{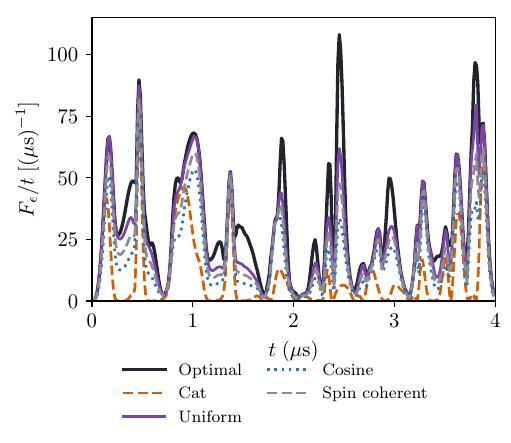}
    \caption{Comparison of the $F_\epsilon(t)/t$ figure of merit for different states under pulse-controlled evolution, with $\epsilon=1.0$.}
    \label{fig:ECHO_QFIovert}
\end{figure}

In Fig.~\ref{fig:ECHO_QFIovert}, we show the $F_\epsilon(t)/t$ figure of merit for the pulse-controlled evolution described by Eq.~\eqref{eq:ECHO_time_evolution}. In contrast to the freely evolving dynamics shown in Fig.~\ref{fig:FID_QFIovert}, several peaks are observed, with the maximal value reached at $t\approx 2.5\,\mu\mathrm{s}$ ($\tau\approx0.62\,\mu\mathrm{s}$). Interestingly, the cat state does not maximize the value of $F_\epsilon(t)/t$. Moreover, the uniform state yields larger values of the figure of merit than the other probe states shown in Fig.~\ref{fig:ECHO_QFIovert}, making it the best-performing fixed probe among those considered.

\section{Conclusions}

In this paper we have studied the problem of estimating parameters encoded in the dephasing of a qudit.
For general dephasing channels, we have derived results on the form and symmetries of optimal probe states and shown that the QFI is concave in the probe populations.
We have also simplified the general SDP for evaluating the channel QFI to a second-order cone program, which gives both the optimal precision and identifies an optimal probe.

We have then considered the case of Gaussian phase diffusion generated by $J_z$.
We have determined the exact range of dephasing strengths for which spin cat states are globally optimal for $J=1$ and $J=3/2$, and derived upper bounds on this range for larger spins.
Our numerical comparison shows that a simple interpolation between cat and uniform probes, optimized for each dephasing strength, is close to optimal over the range considered.
For stronger dephasing, where cat states are no longer optimal, the uniform probes considered in~\cite{Dey2025a} perform near optimally.
This behavior depends on the spectrum of the noise generator: for Gaussian phase diffusion generated by $J_z^2$, the spin cat is insensitive to the noise, while spreading the probe population over all available levels can reduce the precision.

Finally, we have considered a solid-state spin coupled to a nuclear-spin environment, with the dynamics simulated using the CCE technique.
The comparison between free and pulse-controlled evolution shows that the choice of probe strongly depends on the specifics of the dynamics.
For the pulse sequence considered, coherence revivals are accompanied by several peaks in the QFI rate, and we have found that the uniform state performs better than the other fixed probes considered.

Recent experiments have demonstrated the preparation and control of spin cat states in the spin-$7/2$ nucleus of an antimony atom in silicon~\cite{Yu2025c}, providing another setting in which our results could be used to characterize dephasing.
Beyond solid-state spins, neutral-atom qudits could also provide a platform for applying the methods in this paper.
For example, coherent control and interferometry have been demonstrated with the nuclear-spin manifold of $^{87}\mathrm{Sr}$, using tensor light shifts to selectively address transitions~\cite{Ahmed2025a}.
When fluctuations mainly randomize the phases between magnetic sublevels, with negligible population transfer, linear Zeeman shifts and tensor light shifts give the linear and quadratic dephasing generators considered in Sec.~\ref{sec:Gaussian-phase-diffusion}.
Our results could therefore guide the choice of atomic superpositions for characterizing magnetic-field and light-shift fluctuations.
This characterization may be useful for calibrating qudit control and gates, and for assessing residual dephasing in trapping schemes such as the proposed bichromatic tweezers for the metastable $^{3}P_2$ manifold of $^{87}\mathrm{Sr}$~\cite{Carrillo2026}.
Our framework could also be applied to dephasing in spin cat qubit encodings, such as those recently implemented in $^{173}\mathrm{Yb}$~\cite{Kusano2026}.

Quantifying the achievable gains with experimentally feasible preparation and readout operations is a natural next step.
The time required by these operations should also be included when comparing the QFI rate of different probes.
A Bayesian formulation would be useful to account for prior uncertainty about the noise strength and to assess the performance with finite measurement data~\cite{Rubio2019a,Mukhopadhyay2025b}.

Another extension of this work would be to estimate several properties of the dephasing process simultaneously, such as its mean phase and diffusion strength, or the strengths of different diagonal noise couplings.
Joint phase and diffusion estimation is a paradigmatic example in which tradeoffs arise both in the probe state and in the measurement~\cite{Knysh2013,Vidrighin2014,Ragy2016,Szczykulska2017,Jayakumar2024}.
A precise characterization of dephasing is also useful for designing error-correction strategies for spin qudits~\cite{Petiziol2021,Mezzadri2024}.
However, when dephasing itself carries the parameter of interest, correcting its effect can remove useful information from the final probe state.
Understanding how this information can instead be retained in the error-syndrome measurement record, and relating probe optimization to error-corrected channel estimation~\cite{Zhou2021a}, are interesting directions for future work.

The optimization considered in this paper concerns the channel accumulated over a finite interrogation time and does not require Markovian dynamics, provided the channel has the dephasing form of Eq.~\eqref{eq:general_schur_family}.
In Sec.~\ref{sub_sub_sec:CCE} we have considered a prescribed pulse sequence; optimizing the sequence itself, or allowing intermediate measurements and resets, is a further step.
Recent bounds for Markovian noise estimation provide useful benchmarks for these more general strategies~\cite{Gardner2025b,Brady2026}.
Extending such comparisons to temporally correlated noise would be particularly relevant for the spin-bath dynamics considered here.

\section*{Acknowledgments}
The authors thank Cosmo Lupo and Danilo Triggiani for useful discussions.
F.A. and S.C. acknowledge support from the Project PN-RIC PIMIQ (MI1741, B42F26000460005).
This work received financial support from the European Union Next Generation EU, PNRR MUR project PE0000023-NQSTI.

\appendix
\raggedbottom

\section{Second-order cone formulation for optimal estimation of dephasing channels}
\label{app:socp}

For a general dephasing channel $\Lambda_\theta(\rho)=E_\theta\circ\rho$ of the form in Eq.~\eqref{eq:general_schur_family}, the standard optimization to evaluate the channel QFI is simplified because every Kraus operator is diagonal.

\subsection{Purification formula and norm constraints}
Using the notation of Eq.~\eqref{eq:prelim_alpha}, the purification formula~\cite{Escher2012a} gives, at regular parameter values,
\begin{equation}
    F_\theta(p)=4\inf_{h=h^\dagger}\sum_m p_m[\alpha_\theta(h)]_{mm}.
    \label{eq:population_variational_qfi}
\end{equation}
Eq.~\eqref{eq:population_variational_qfi} also proves concavity, since its right-hand side is an infimum of affine functions of $p$.

For a dephasing channel, consider a Kraus representation $\Lambda_\theta(\rho)=\sum_k K_k\rho K_k^\dagger$ in which every Kraus operator is diagonal
\begin{equation}
    K_k=\sum_m c_{km}\ket m\bra m,
    \qquad c_m=(c_{1m},\ldots,c_{rm})^{\mathsf T},
\end{equation}
where $r$ is the number of Kraus operators.
Then $E_{mn}=\sum_k c_{km}c_{kn}^*$ and, since $E_{mm}=1$, we have that $\|c_m\|_2=1$. 
A Hermitian matrix $h$ describes the freedom to change the Kraus representation locally. For each level $m$, define
\begin{equation}
    v_m(h)=\dot c_m-ihc_m.
\end{equation}
Because the Kraus operators are diagonal, $\alpha_\theta(h)$ is also diagonal with $[\alpha_\theta(h)]_{mm}=\|v_m(h)\|_2^2$. 
Considering Eq.~\eqref{eq:population_variational_qfi}, for the channel QFI we have
\begin{equation}
    \frac{F_Q[\Lambda_\theta]}{4}=\max_{p}\min_{h=h^\dagger}\sum_m p_m \|v_m(h)\|_2^2.
\end{equation}
The dependence on $p$ is linear, while each $\|v_m(h)\|_2^2$ is convex in $h$ because $v_m(h)$ is an affine function of $h$.
The minimax theorem therefore allows the order of the two optimizations to be exchanged. Maximizing a weighted average over all probability distributions $p$ then selects the largest of the $\|v_m(h)\|_2^2$, giving
\begin{equation}
    \frac{F_Q[\Lambda_\theta]}{4}=\min_{h=h^\dagger}\max_m \|v_m(h)\|_2^2.
    \label{eq:schur_minimax}
\end{equation}
By Sec.~\ref{subsec:probe_populations}, this optimum is attainable without an ancilla. 
Introducing an auxiliary variable $t$, Eq.~\eqref{eq:schur_minimax} is equivalently expressed through the constraints $\|v_m(h)\|_2\le t$ for every $m$.
Minimizing $t$ gives the SOCP in Eq.~\eqref{eq:schur_socp}, and its optimal value $t^\star$ determines $F_Q[\Lambda_\theta]=4{t^\star}^2$.
For complex coefficients, stacking the real and imaginary parts of $v_m$ yields ordinary real norm constraints.

The general SDP instead imposes a matrix positivity constraint. Eq.~\eqref{eq:schur_socp} makes its dephasing structure explicit using one norm constraint per level.
This is a specialization of the established channel-QFI framework of~\cite{Zhou2021a}, Eqs.~(5)--(6) and Appendix~F.
Their Appendix~G already notes a quadratic simplification for a lossy interferometer.
Here we give the explicit cone formulation for dephasing channels.

For real, reflection-invariant channels, the optimization over $h$ can be simplified further as described in Sec.~\ref{app:real_reflection_gauges} below.

\subsection{Optimal probe populations from the dual variables}
The scalar dual multipliers associated with the norm constraints can be identified with the populations of an optimal probe state. Let $\mu_m$ be the scalar dual variable for the $m$th norm constraint in Eq.~\eqref{eq:schur_socp}. 
At the dual optimum,
\begin{equation}
    p_m=\mu_m^\star,\qquad
    \mu_m^\star\geq0,\qquad\sum_m\mu_m^\star=1
    \label{eq:socp_populations}
\end{equation}
gives optimal probe populations. Thus one may directly prepare
\begin{equation}
    \ket{\psi_\star}=\sum_m\sqrt{p_m}\ket m,
\end{equation}
a pure state of the probe alone, obtained from the same optimization. For a general channel, the optimal input is a purification of a mixed state of the system, and finding it requires an additional calculation~\cite{Zhou2021a}.

To prove Eq.~\eqref{eq:socp_populations}, write the norm constraints as 
 \begin{equation}
    \|v_m(h)\|_2-t\leq0. 
 \end{equation}
 The Lagrangian is 
 \begin{equation} 
    L(t,h,\mu) = t+\sum_m\mu_m\left(\|v_m(h)\|_2-t\right), \qquad \mu_m\geq0, 
 \label{eq:population_lagrangian} 
 \end{equation}
 where $\mu=(\mu_1,\ldots,\mu_d)^{\mathsf T}$. 
 The chosen nonnegativity of the multipliers implies that $L(t,h,\mu)\leq t$ for every primal feasible pair $(t,h)$ and every $\mu\geq0$. 
By definition, the dual function is $\inf_{t,h=h^\dagger}L(t,h,\mu)$
and, since $t\in\mathbb R$ is unrestricted, finiteness of the infimum over $t$ requires $\frac{\partial L}{\partial t}=0$. Then $\sum_m\mu_m=1$ and the dual multipliers form a probability vector. Let $(t^\star,h^\star,\mu^\star)$ be a primal-dual optimal point.
Define 
    \begin{equation} Q_{\mu^\star}(h) = \sum_m\mu_m^\star\|v_m(h)\|_2^2. 
    \label{eq:weighted_squared_norm}
\end{equation}
Slater’s condition is verified because, for any $h$, one may choose $t>\max_m \|v_m(h)\|_2$, hence strong duality holds.
Then the complementary slackness conditions
\begin{equation}
    \mu_m^\star \left( \|v_m(h^\star)\|_2-t^\star \right) =0.   \label{eq:compl_slackness}
\end{equation} 
are verified.
Therefore, we have
\begin{equation}
    \|v_m(h^\star)\|_2=t^\star  \qquad  \mu_m^\star>0
    \label{eq:norm_active_constraints}
\end{equation}
and $Q_{\mu^\star}(h^\star)={t^\star}^2$.
Suppose first that $t^\star>0$. Stationarity of the Lagrangian with respect to the Hermitian variable $h$ gives 
\begin{equation} 
    0 = \nabla_h L(t^\star,h^\star,\mu^\star) = \sum_{m:\mu_m^\star>0}\mu_m^\star \nabla_h\|v_m(h^\star)\|_2.
    \label{eq:lagrangian_h_stationarity} 
\end{equation} 
The sum is restricted to $m$ such that $\mu_m^\star>0$, since if $\mu_m^\star=0$, Eq.~\eqref{eq:norm_active_constraints} need not hold.
In that case, we may obtain $\|v_m(h^\star)\|_2=0$, so that the function $\|v_m(h)\|_2$ is not differentiable at $h^\star$.
Note that Eq.~\eqref{eq:compl_slackness} shows that a level can have positive optimal population only if its norm constraint is active. Activity, however, is only a necessary condition because arbitrary probability weights on the active levels need not satisfy the stationarity condition \eqref{eq:lagrangian_h_stationarity}.
Moreover, 
\begin{equation}
    \nabla_h Q_{\mu^\star}(h^\star) = 2t^\star \sum_{m:\mu_m^\star>0}\mu_m^\star \nabla_h\|v_m(h^\star)\|_2 =0  
    \label{eq:Q_stationarity}
\end{equation} 
where, in the first equality computed the gradient and used  Eq.~\eqref{eq:norm_active_constraints}, while in the second one we exploited the relation in Eq.~\eqref{eq:lagrangian_h_stationarity}. 
Since the function $Q_{\mu^\star}(h)$ is convex, Eq.~\eqref{eq:Q_stationarity} therefore implies that $h^\star$ is a global minimizer.
By considering Eq.~\eqref{eq:population_variational_qfi} we see that
\begin{equation}
F_\theta(\mu^\star)=4\inf_{h=h^\dagger}Q_{\mu^\star}(h)=4{t^\star}^2=F_Q[\Lambda_\theta]
\end{equation}
Thus $p_m^\star=\mu_m^\star$ is an optimal probe population because it attains the channel QFI.
If $t^\star=0$, then the channel QFI vanishes.
Since the QFI is nonnegative, every probe has zero QFI and is therefore optimal.  

If reflection symmetry is used to restrict $h$, the recovered populations should also be symmetrized, $p_m=(\mu_m+\mu_{d+1-m})/2$. If only one cone per reflection pair is retained in the primal problem, its dual weight is the total population of the pair and is shared equally between its two levels; an unpaired central level receives its own weight.

To conclude, we report the dual problem that can be solved to get both $h^\star$ and the optimal population vector:
\begin{equation}
\begin{aligned} 
& \underset{\mu}{\operatorname{max}} \quad \underset{h=h^\dagger}{\operatorname{inf}} \quad \sum_m \mu_m \left\|v_m(h)\right\|_2, \\ & \text{subject to}\quad \mu_m\geq 0, \quad \sum_m \mu_m=1.
\end{aligned}
\end{equation}

\subsection{Real Kraus operators and reflection symmetry}
\label{app:real_reflection_gauges}
For real $c_m$ and $\dot c_m$, decompose $h=S+iA$, where $S=S^{\mathsf T}$ and $A=-A^{\mathsf T}$ are real. Then
\begin{equation}
    q_m(h)=\|\dot c_m+Ac_m\|_2^2+\|Sc_m\|_2^2.
\end{equation}
Setting $S=0$ cannot increase any constraint, so it suffices to optimize over the real antisymmetric matrix $A$.

Suppose also that $E$ commutes with the reversal $U$ throughout the parameter interval. Choose a fixed real orthogonal matrix $O$ such that $OUO^{\mathsf T}=\Sigma=\operatorname{diag}(\sigma_k)$, with $\sigma_k=\pm1$, and take the real Kraus coefficient matrix $C=O\sqrt E$. Its columns $c_m$ satisfy $C^{\mathsf T}C=E$ and, writing $m'=d+1-m$,
\begin{equation}
    c_{m'}=\Sigma c_m,\qquad \dot c_{m'}=\Sigma\dot c_m.
\end{equation}
Thus $q_m(i\Sigma A\Sigma)=q_{m'}(iA)$. Since every $q_m(iA)=\|\dot c_m+Ac_m\|_2^2$ is convex in $A$, the averaged matrix $\bar A=(A+\Sigma A\Sigma)/2$ obeys
\begin{equation}
    \max_m q_m(i\bar A)\leq\max_m q_m(iA).
\end{equation}
It also satisfies $\bar A_{kl}=0$ whenever $\sigma_k\ne\sigma_l$. The optimization can therefore be restricted to two antisymmetric parity blocks, of dimensions $\lceil d/2\rceil$ and $\lfloor d/2\rfloor$, reducing the number of numerical variables. 

\section{Mapping between models: multiple qubits, single spin and two-mode photonic states}
\label{app:spin_mapping}

Consider $N$ qubits coupled identically to the same fluctuating field, so that they acquire the same random phase~\cite{Rossi2015,Dey2025a}.
Writing $S_z=\frac12\sum_{k=1}^N\sigma_z^{(k)}$, the collective dephasing channel is
\begin{equation}
    \Lambda_\theta^{(N)}(\rho)=\int_{\mathbb R}d\phi\,
    w_\theta(\phi)e^{-i\phi S_z}\rho e^{i\phi S_z}.
    \label{eq:collective_qubit_channel}
\end{equation}
The fully symmetric subspace has dimension $N+1$ and carries the spin-$J=N/2$ representation.
Its normalized Dicke state $\ket{D_m^{(N)}}$, containing $J+m$ spins up, is identified with the spin state $\ket m$, for $m=-J,\ldots,J$.
Since $S_z\ket{D_m^{(N)}}=m\ket{D_m^{(N)}}$, the restriction of Eq.~\eqref{eq:collective_qubit_channel} to this subspace is exactly the spin dephasing channel of Sec.~\ref{subsec:dephasing_preliminaries}, with
\begin{equation}
    (E_\theta)_{mn}=\int_{\mathbb R}d\phi\,
    w_\theta(\phi)e^{-i(m-n)\phi}.
\end{equation}
In particular, a zero-mean Gaussian phase of variance $2\Delta$ gives $(E_\Delta)_{mn}=e^{-\Delta(m-n)^2}$.

Restricting the input to the fully symmetric subspace does not reduce the optimal QFI.
By convexity, it suffices to consider pure inputs.
Let $\Pi_m$ project onto the full eigenspace of $S_z$ with eigenvalue $m$.
Any parameter-independent pure probe can be written as
\begin{equation}
    \ket\psi=\sum_m\sqrt{p_m}\ket{v_m},\qquad
    p_m=\bra\psi\Pi_m\ket\psi,
\end{equation}
where $\ket{v_m}$ is a normalized vector in that eigenspace and can be chosen arbitrarily when $p_m=0$.
Within each eigenspace, choose a unitary mapping $\ket{v_m}$ to $\ket{D_m^{(N)}}$.
Their direct sum defines a parameter-independent unitary $W$ satisfying $[W,S_z]=0$ and
\begin{equation}
    W\ket\psi=\sum_m\sqrt{p_m}\ket{D_m^{(N)}}
    \equiv\ket{\psi_{\mathrm{sym}}}.
\end{equation}
Consequently, from Eq.~\eqref{eq:collective_qubit_channel}, we get
\begin{equation}
    \Lambda_\theta^{(N)}(\ket{\psi_{\mathrm{sym}}}\bra{\psi_{\mathrm{sym}}})
    =W\Lambda_\theta^{(N)}(\ket\psi\bra\psi)W^\dagger.
    \label{eq:symmetric_qubit_equivalence}
\end{equation}
By unitary invariance, the two probes $\ket\psi$ and $\ket{\psi_{\mathrm{sym}}}$ have the same QFI, thus every pure probe has a fully symmetric counterpart with identical precision.
This argument also applies when the random phase is generated by a fixed function of $S_z$, such as $S_z^2$.
The reduction relies on collective coupling; independent or general spatially correlated dephasing is not described by the same spin channel.

For two bosonic modes $a,b$ with fixed total photon number $N$, the Schwinger representation gives the same spin-$J=N/2$ system~\cite{Lee2002,Demkowicz-Dobrzanski2015a}:
\begin{equation}
    J_z=\frac{a^\dagger a-b^\dagger b}{2},\qquad
    \ket m\longleftrightarrow\ket{J+m}_a\ket{J-m}_b.
    \label{eq:spin_fock_mapping}
\end{equation}
A relative phase shift is generated by $J_z$, so random fluctuations of this phase produce the dephasing channel considered above~\cite{Knysh2013}.
Fixed beam splitters before and after the phase encoding can be absorbed into the probe preparation and measurement, respectively, when optimising the QFI.
Under these identifications, the qubit GHZ state $(\ket{\uparrow}^{\otimes N}+\ket{\downarrow}^{\otimes N})/\sqrt2$, the spin cat $(\ket J+\ket{-J})/\sqrt2$, and the photonic NOON state $(\ket N_a\ket0_b+\ket0_a\ket N_b)/\sqrt2$ are equivalent probes.
The same correspondence maps a uniform Dicke superposition to a uniform superposition of spin levels and of the $N+1$ two-mode Fock states.

\section{QFI calculations and cat-optimality bounds for Gaussian diffusion}
\label{app:gaussian_derivations}
We give the calculations underlying Sec.~\ref{subsec:cat_optimality}. Throughout this appendix $u=e^{-2\Delta}$ and dots denote derivatives with respect to $\Delta$, with the probe populations held fixed. For blocks with parameter-independent orthogonal supports,
\begin{equation}
    F_Q\!\left[\bigoplus_k\tau_k\rho_k\right]
    =\sum_k\left[\frac{\dot\tau_k^2}{\tau_k}+\tau_k F_Q[\rho_k]\right].
    \label{eq:QFI_orthosupp}
\end{equation}
For a full-rank qubit state $\rho=(\mathbbm{1}+\bm r\cdot\bm\sigma)/2$, we use
\begin{equation}
    F_Q[\rho]=\|\dot{\bm r}\|^2+
    \frac{(\bm r\cdot\dot{\bm r})^2}{1-\|\bm r\|^2}.
    \label{eq:QFI_qubit}
\end{equation}
The formulas below were obtained for $0\le\epsilon\le1$ and $0<u<1$.

\subsection{Spin $1$}
\label{app:spin_one_qfi}
For the probe in Eq.~\eqref{eq:spin_one_probe}, use the basis $\{\ket{s},\ket{0},\ket{a}\}$, where $\ket{s}=(\ket{-1}+\ket{1})/\sqrt2$ and $\ket{a}=(\ket{-1}-\ket{1})/\sqrt2$. The output is $\tau\rho_q\oplus\lambda$, with
\begin{align}
    \rho_q&=\frac1\tau
    \begin{pmatrix}
        (1-\epsilon)(1+u^2)/2 & \sqrt{\epsilon(1-\epsilon)u}\\
        \sqrt{\epsilon(1-\epsilon)u} & \epsilon
    \end{pmatrix},\\
    \lambda&=\frac{(1-\epsilon)(1-u^2)}2,\qquad
    \tau=\frac{1+\epsilon+(1-\epsilon)u^2}2.
\end{align}
Eqs.~\eqref{eq:QFI_orthosupp}--\eqref{eq:QFI_qubit} give
\begin{equation}
    f_1(\epsilon)=
    \frac{8u(1-\epsilon)[2u^3-\epsilon(2u^3+u^2-2u-1)]}
    {(1-u^2)[1+u^2+\epsilon(1-u^2)]}.
    \label{eq:spin_one_full_qfi}
\end{equation}
At $\epsilon=0$ this is $16u^4/(1-u^4)$, the cat QFI. Differentiating gives Eq.~\eqref{eq:spin_one_endpoint}. The full derivative is
\begin{equation}
    f'_1(\epsilon)=
    -\frac{8u[A(u)\epsilon^2+B(u)\epsilon+C(u)]}
    {(1-u^2)[1+u^2+\epsilon(1-u^2)]^2},
    \label{eq:QFI_spin_one_derivative}
\end{equation}
where $C$ is defined in Eq.~\eqref{eq:spin_one_threshold_polynomial} and
\begin{align}
    A(u)&=(1-u^2)^2(1+2u),\\
    B(u)&=2(1-u^2)(1+2u)(1+u^2).
\end{align}
From the full derivative, we also obtain the optimal central population when the cat state is not optimal. The function $f_1$ is strictly concave, since
\begin{equation}
    f''_1(\epsilon)=
    -\frac{32u(1+u)^2}{[1+u^2+\epsilon(1-u^2)]^3}<0.
\end{equation}
For $\Delta>\Delta_1$, $f'_{1}(0)>0$ and $f_1(1)=0<f_1(0)$, so strict concavity gives a unique interior maximizer. Solving Eq.~\eqref{eq:QFI_spin_one_derivative} yields
\begin{equation}
    \epsilon_\star(u)=
    \frac{\sqrt2(1+u)\sqrt{1+2u}-(1+2u)(1+u^2)}
    {(1-u^2)(1+2u)}.
    \label{eq:spin_one_optimal_population}
\end{equation}
For $0<\Delta\leq\Delta_1$, the optimum is instead $\epsilon_\star=0$.

\subsection{Spin $3/2$}
\label{app:spin_three_halves_qfi}
For Eq.~\eqref{eq:spin_three_halves_probe}, define the outer and inner parity states $\ket{\sigma_o}=(\ket{-3/2}+\sigma\ket{3/2})/\sqrt2$ and $\ket{\sigma_i}=(\ket{-1/2}+\sigma\ket{1/2})/\sqrt2$, with $\sigma=\pm1$. The output has two qubit blocks, $\tau_+\rho_+\oplus\tau_-\rho_-$, where
\begin{equation}
    \tau_\sigma=\frac{(1-\epsilon)(1+\sigma u^{9/2})+
    \epsilon(1+\sigma u^{1/2})}{2}.
    \label{eq:tau-sigma}
\end{equation}
In each parity basis $\{\ket{\sigma_o},\ket{\sigma_i}\}$,
\begingroup\small
\begin{equation}
    \rho_\sigma=\frac1{2\tau_\sigma}
    \begin{pmatrix}
        (1-\epsilon)(1+\sigma u^{9/2}) & \sqrt{\epsilon(1-\epsilon)}(u^{1/2}+\sigma u^2)\\
        \sqrt{\epsilon(1-\epsilon)}(u^{1/2}+\sigma u^2) & \epsilon(1+\sigma u^{1/2})
    \end{pmatrix}.
    \label{eq:B_sigma}
\end{equation}
\endgroup
Using Eqs.~\eqref{eq:QFI_orthosupp}--\eqref{eq:QFI_qubit} gives
\begin{equation}
    f_{3/2}(\epsilon)=r(u)
    \frac{\sum_{k=0}^3p_k(u)\epsilon^k}{\sum_{k=0}^2q_k(u)\epsilon^k},
    \label{eq:spin_three_halves_full_qfi}
\end{equation}
with the coefficients listed below.
\begin{widetext} 
\begin{align} 
r(u) ={}&\frac{u}{(u+1)(u^3-1)} \\[1mm] 
p_0(u) ={}&-81u^{11}-162u^{10}-162u^9-81u^8 \\ 
p_1(u) ={}&242u^{11}+479u^{10}+491u^9+288u^8 -36u^7-162u^6-189u^5  -117u^4-72u^3-35u^2-20u-5 \\ 
p_2(u) ={}&-241u^{11}-472u^{10}-496u^9-333u^8 +70u^7+310u^6+361u^5 +243u^4+117u^3+40u^2+13u+4 \\ 
p_3(u) ={}&80u^{11}+155u^{10}+167u^9+126u^8 -34u^7-148u^6-172u^5 -126u^4-46u^3-7u^2+5u \\ 
q_0(u) ={}&u^8+u^7+u^6+u^5+u^4+u^3+u^2+u+1 \\ 
q_1(u) ={}&-2u^8-2u^7-2u^6-2u^5 \\ 
q_2(u) ={}&u^8+u^7+u^6+u^5-u^4-u^3-u^2-u \\[1mm] 
t_0(u) ={}&80u^{19}+235u^{18}+402u^{17}+528u^{16} +654u^{15}+816u^{14}+951u^{13} +996u^{12}+924u^{11}+647u^{10} \nonumber\\ &\quad +148u^9 -348u^8-636u^7-600u^6 -438u^5-249u^4-132u^3-60u^2-25u-5
\end{align}
\end{widetext}
In particular, $t_0=p_1q_0-p_0q_1$. At $\epsilon=0$, Eq.~\eqref{eq:spin_three_halves_full_qfi} gives $81u^9/(1-u^9)$, the spin-cat QFI, and $q_0(u)>0$. Hence its derivative is Eq.~\eqref{eq:spin_three_halves_endpoint}. Since the QFI is concave in the populations, the sign of this derivative determines whether the cat state is globally optimal.

\subsection{Bounds from spin kitten states}

\label{app:larger_spin_bounds}

The first two bounds in Eq.~\eqref{eq:cat_region_bounds} are obtained by restricting the probe to three or four levels, which reproduces the spin-one and spin-$3/2$ problems, as discussed in Sec.~\ref{subsec:cat_optimality}. To bound the remaining half-integer spins, compare Eq.~\eqref{eq:gaussian_cat_qfi} for $m<J$ and for $m=J$. With $s=m^2/J^2\in(0,1)$ and $\delta=8J^2\Delta$, the spin kitten is better when
\begin{equation}
    s^2(e^\delta-1)>e^{s\delta}-1.
\end{equation}
The difference of these two sides has the expansion
\begin{equation}
    (s^2-s)\delta+\sum_{n=3}^\infty
    \frac{(s^2-s^n)\delta^n}{n!}.
\end{equation}
After division by $\delta>0$, the constant term is negative and every remaining coefficient is positive. The expression is strictly increasing from a negative value to infinity, so it has exactly one positive root $\delta^*(s)$, defined by
\begin{equation}
    s^2(e^{\delta^*(s)}-1)=e^{s\delta^*(s)}-1.
    \label{eq:condition}
\end{equation}
Consequently,
\begin{equation}
    \Delta_{\mathrm{cat}}(J)\leq\frac{\delta^*(s)}{8J^2}.
    \label{eq:implicit_bound}
\end{equation}
To obtain a constant independent of $J$, set $g(t)=(e^t-1)/t^2$ and $x=\delta^*(s)$. Eq.~\eqref{eq:condition} then reads
\begin{equation}
    g(x)=g(sx).
    \label{eq:auxiliary_cond}
\end{equation}
The function $g$ decreases up to its unique positive stationary point $t_*\approx1.5936$ and increases thereafter. Since $sx<x$, from Eq.~\eqref{eq:auxiliary_cond} one has $sx<t_*<x$, then $g'(sx)<0$ and $g'(x)>0$. Implicit differentiation of Eq.~\eqref{eq:auxiliary_cond} with respect to $s$ therefore gives
\begin{equation}
    \frac{dx}{ds}=\frac{xg'(sx)}{g'(x)-sg'(sx)}<0.
\end{equation}
For half-integer $J\geq5/2$, choose $m=J-1$, so $s=(1-1/J)^2$ increases with $J$. The largest value of $\delta^*(s)$ is then obtained at $J=5/2$, where $s=9/25$ and
\begin{equation}
    \delta^*(9/25)\approx2.514731,\qquad
    \Delta_{5/2}=\frac{\delta^*(9/25)}8\approx0.314341.
\end{equation}
This proves the bound
\begin{equation}
    \Delta_{\mathrm{cat}}(J)\leq\frac{\Delta_{5/2}}{J^2},
    \qquad J=5/2,7/2,9/2,\ldots,
    \label{eq:explicit_bound}
\end{equation}
which is replaced by the tighter bound obtained from the spin-$3/2$ problem when $J=3n/2$ with odd $n$.

\bibliography{2026DephasingEstimation}
\end{document}